# Proton dynamics in water confined at the interface of the graphene-MXene heterostructure

Lihua Xu[1] and De-en Jiang[1,2,*]

[1]Department of Chemical and Environmental Engineering, University of California, Riverside, California 92521, USA

[2]Department of Chemistry, University of California, Riverside, California 92521, USA

*E-mail: djiang@ucr.edu

## Abstract

Heterostructures of 2D materials offer a fertile ground to study ion transport and charge storage. Here we employ *ab initio* molecular dynamics to examine the proton-transfer/diffusion and redox behavior in a water layer confined in the graphene-$Ti_3C_2O_2$ heterostructure. We find that in comparison with the similar interface of water confined between $Ti_3C_2O_2$ layers, proton redox rate in the dissimilar interface of graphene-$Ti_3C_2O_2$ is much higher, owning to the very different interfacial structure as well as the interfacial electric field induced by an electron transfer in the latter. Water molecules in the dissimilar interface of the graphene-$Ti_3C_2O_2$ heterostructure form a denser hydrogen-bond network with a preferred orientation of water molecules, leading to an increase of proton mobility with proton concentration in the graphene-$Ti_3C_2O_2$ interface. As the proton concentration further increases, proton mobility deceases, due to increasingly more frequent surface redox events that slow down proton mobility due to binding with surface O atoms. Our work provides important insights into how the dissimilar interface and their associated interfacial structure and properties impact proton transfer and redox in the confined space.

# I. Introduction

MXenes, a two-dimensional (2D) transition metal carbides and nitrides, first discovered in 2011,[1,2] have emerged as a versatile material for various applications, including energy storage,[3–7] membranes,[8,9] electronics,[10–12] sensors,[13–18] and catalysts.[19–22] The high electrical conductivity and high volumetric capacitance make MXenes a promising electrode material for energy storage.[23] Previous experimental studies reported that proton-involved reversible surface redox reaction was the key to the pseudocapacitive behavior of MXene electrodes in the aqueous $H_2SO_4$ electrolyte,[24–25] while a recent computational study showed that the thickness of the water layer impacts the rates of the surface redox reaction and the proton transport in the MXene-confined water.[26]

Hybridizing MXenes with carbon-based materials (graphene nano-flakes, carbon nanotubes, etc.), metal oxides, or polymers[27–33] have been found to further enhance the performance of energy storge. For example, graphene/$Ti_2CT_x$@polyaniline composite[34] and MXene/reduced-graphene-oxide electrodes[6] exhibit higher capacitances and excellent cycling stability in the $H_2SO_4$ electrolyte. However, how the interface of MXene/graphene impacts proton surface redox and proton transport in the electrolyte is still unclear. Recent computational studies have shed light on different heterostructures and their dissimilar interfaces, focusing on the electronic structure of the electrode itself.[35,36,37]

The experimental studies of MXene/graphene composites and the computational studies of various heterostructures prompted us to seek the fundamental understanding of the capacitive energy-storage mechanism of electrolytes at the dissimilar interface, especially proton redox and transport. To this end, here we explore the effects of the interfacial properties and configurations on proton redox and dynamics in a water layer confined between graphene and MXene layers via *ab initio* molecular dynamics (AIMD). The proton was chosen because of its fast surface redox reaction and transport in water during the pseudocapacitive energy-storage process in MXenes that is accessible to the timescale of AIMD.[38] We chose $Ti_3C_2O_2$ as a representative MXene, whose -O termination will be partially converted to -OH in an acid electrolyte after binding with a proton. Below we first elaborate on our computational methods.

## II. Computational methods

The Vienna *ab initio* Simulation Package (VASP[39,40]) was used for both structure optimization and *ab initio* molecule dynamics (AIMD) simulations based on density-functional theory (DFT) with periodic boundary conditions. The electron-ion interactions were described by the projector augmented-wave methods[41,42], while electron exchange-correlation by the Perdew-Burke-Ernzerhof (PBE[43]) functional form of the generalized-gradient approximation (GGA). The kinetic energy cutoff of 500 eV was used for the plane-wave basis set. Grimme's DFT-D3 method with the Becke-Jonson damping[44] was used for the van der Waals (vdW) interactions.[45,46]

To model the dissimilar interface, we used a monoclinic supercell for both graphene-MXene (a 4×4 supercell of $Ti_3C_2O_2$ matched to a 5×5 supercell of graphene with a lattice mismatch of ~1.3%). For comparison, we also examined the MXene-MXene interface (a 4×4 supercell of $Ti_3C_2O_2$). To avoid dipoles along the c direction, symmetric cells are used to model water/proton confined in graphene-MXene and MXene-MXene interfaces (Fig. 1). For convenience, MO_MO and G_MO_G denote $Ti_3C_2O_2$-$Ti_3C_2O_2$ and graphene-$Ti_3C_2O_2$-graphene systems, respectively. Each interface contains a single water layer of 12 water molecules and 0 to 3 protons; the whole simulation cell is charge neutral. We use "$n$p" to denote the number of protons being intercalated, e.g., G_1p_MO_1p_G means 1 proton per interface of graphene-$Ti_3C_2O_2$.

The super cells were structurally optimized with convergence criteria of 0.02 eV/Å in force and $10^{-5}$ eV in energy; the Brillouin zone was sampled by the 3×3×1 Monkhorst-Pack grid.[47] Coordinates for the optimized structures are provided in the Supplementary Material and the optimized c-lattice parameters, corresponding to the supercells in Fig. 1, are shown in Table 1. With the optimized structures as input, AIMD simulations were performed via the canonical ensemble (NVT) with the Nosé-thermostat at 300 K for 20 ps at 1 fs time-step[48–50] on the optimized structures with Γ-point only for the k-point sampling. The last 15 ps of trajectories were used for the analysis. Proton movement was tracked by monitoring the hydronium O atoms; the diffusion coefficient was calculated from the Einstein relation and the time-dependent mean square displacement. See supplementary material for details.

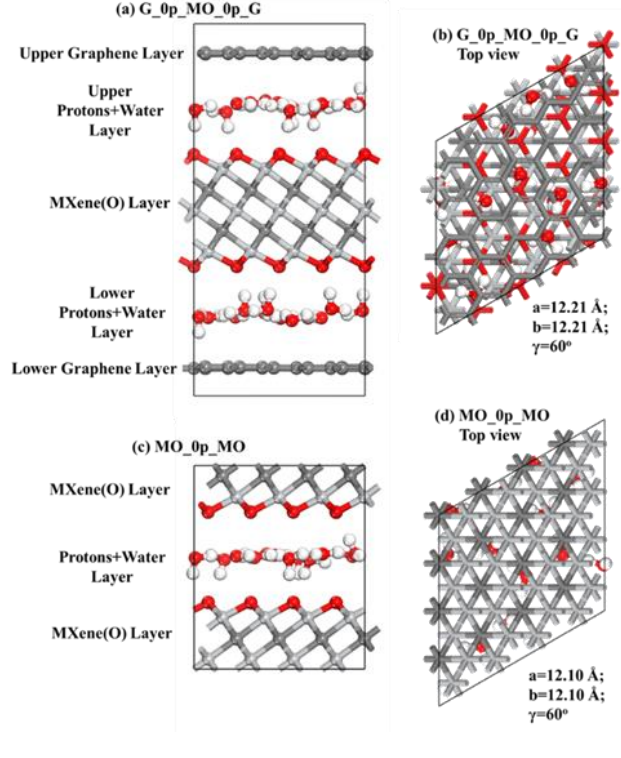

**Fig. 1.** Structure models: (a) side and (b) top views of water/proton confined in graphene-$Ti_3C_2O_2$-graphene (denoted as G_MO_G); (c) side and (d) top views of water/proton confined in $Ti_3C_2O_2$-$Ti_3C_2O_2$ (denoted as MO_MO). In the supercells, each interface has 12 molecules and $n$ protons ($n$p, n from 0 to 3).

**Table 1.** Optimized c-axis lattice parameters for different MO_MO and G_MO_G systems.

| System | c (Å) | System | c (Å) |
|---|---|---|---|
| G_1p_MO_1p_G | 23.0 | MO_1p_MO | 12.5 |
| G_2p_MO_2p_G | 23.0 | MO_2p_MO | 13.0 |
| G_3p_MO_3p_G | 23.0 | MO_3p_MO | 13.0 |

## III. Results and discussion

### A. Proton transport in water versus proton surface redox reaction

The goal of the present work is to reveal how the dissimilar interface of $Ti_3C_2O_2$-graphene impacts proton redox chemistry and transport. To simplify the problem, we used a monolayer of water but varying proton concentrations. Both in-water proton transfer and reversible surface redox processes of proton can take place at the $Ti_3C_2O_2$-graphene interface where protons in hydronium ions ($H_3O^+$) transfer to O atoms on $Ti_3C_2O_2$ surface to form hydroxyl groups (-OH) which after certain time interval can release protons back to water molecules. Graphene, on the other hand, is inert, with respect to proton redox chemistry. From the AIMD trajectories, we tracked the events of in-water proton transfer vs surface redox reaction. As shown in Fig. 2(a) for the $Ti_3C_2O_2$-graphene interface, in-water proton transfer is much more frequent than the surface redox reactions which, as the concentration of proton increases, become more frequent. Comparing with the $Ti_3C_2O_2$-$Ti_3C_2O_2$ interface [Fig. 2(b)], surface redox reactions are much more frequent at the $Ti_3C_2O_2$-graphene interface. In the case of three protons, the estimated surface redox rate is 634 m/s at the $Ti_3C_2O_2$-graphene interface (G_3p_MO_3p_G), about 9 times that at the $Ti_3C_2O_2$-$Ti_3C_2O_2$ interface (MO_3p_MO). Therefore, both interfacial proton concentration and interface compositions can impact the proton surface-redox behavior. Below we analyze the proton mobility in detail.

**B.     Proton diffusivity and trajectories at the interfaces**

By tracking the O atoms with an extra hydrogen atom such as O in the hydronium ion ($H_3O^+$) or the hydroxyl O on $Ti_3C_2O_2$ surfaces (Fig. S1) and determining their root-mean-square displacements (Fig. S2), we determined the proton diffusion coefficients at the $Ti_3C_2O_2$-graphene and $Ti_3C_2O_2$-$Ti_3C_2O_2$ interfaces. As one can see from **Fig. 3**, proton diffusivity decreases with the increasing proton concentration at the $Ti_3C_2O_2$-$Ti_3C_2O_2$ interface, but shows a maximum at 2 protons per water layer at the $Ti_3C_2O_2$-graphe e interface. Though it is about the same at 1 proton per water layer for both interfaces, proton diffusivity in $Ti_3C_2O_2$-graphene is more than twice that in $Ti_3C_2O_2$-$Ti_3C_2O_2$ at 2 or 3 protons per water layer. In other words, at higher proton concentrations proton diffusivity is much higher at the dissimilar interface than the similar interface. Below we analyze the proton diffusivity in detail in order to understand the two different trends in Fig. 3.

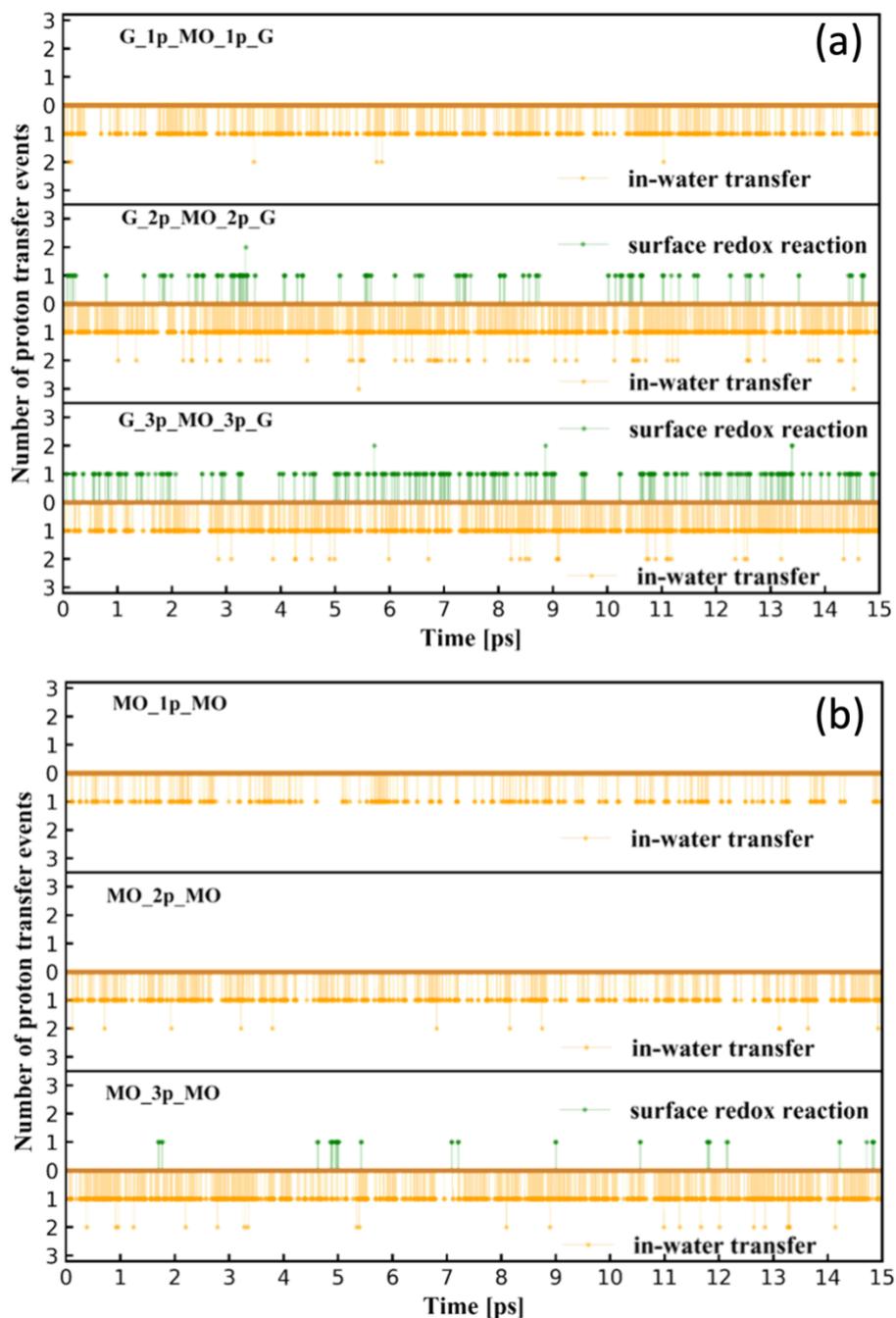

**Fig. 2.** Number of proton surface-redox (green) versus in-water proton transfer (orange) events with time for different numbers of intercalated protons ($n$p) at the interface: (a) graphen-Ti$_3$C$_2$O$_2$-graphene (G_$n$p_MO_$n$p_G) systems; (b) Ti$_3$C$_2$O$_2$-Ti$_3$C$_2$O$_2$ (MO_$n$p_MO) systems.

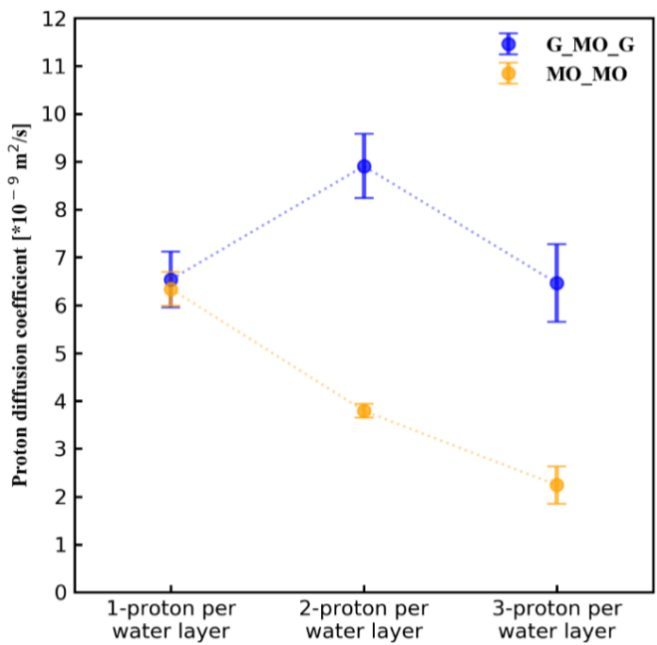

**Fig. 3.** Proton diffusivity at the $Ti_3C_2O_2$-graphene and $Ti_3C_2O_2$-$Ti_3C_2O_2$ interfaces vs different amounts of intercalated protons.

Since proton can be either in $H_3O^+$ or -OH group on $Ti_3C_2O_2$, we have estimated the probabilities of proton in the two states from the AIMD trajectories. As shown in Table 1, proton remains in the water layer almost all the time in the case of 1 or 2 protons in the $Ti_3C_2O_2$-graphene interface. We further tracked the hydronium ions in these two cases (Fig. 4): one can see that the motion of the one hydronium ion is restricted in one region and half of the space is rarely explored [Fig. 4(a)] in the 1p case, but the two hydronium ions now explore almost all the interfacial space [Fig. 4(b)] in the 2p case. In the case of 3p, they roughly split the time equally in water or on the $Ti_3C_2O_2$ surface (Table 1). This can explain the decrease in the overall proton diffusivity, because the proton in -OH group on the $Ti_3C_2O_2$ surface is much less mobile than in water, as evidenced by the localized proton sites and more unexplored space at the interface [Fig. 4(c)].

**Table 2.** Percent of time when protons are in $H_3O^+$ in water layer or as -OH groups on MXene surface for G_MO_G systems.

| Systems | % of time | |
|---|---|---|
| | as $H_3O^+$ | as -OH on $Ti_3C_2O_2$ |
| G_1p_MO_1p_G | 100 | 0 |
| G_2p_MO_2p_G | 96 | 4 |
| G_3p_MO_3p_G | 53 | 47 |

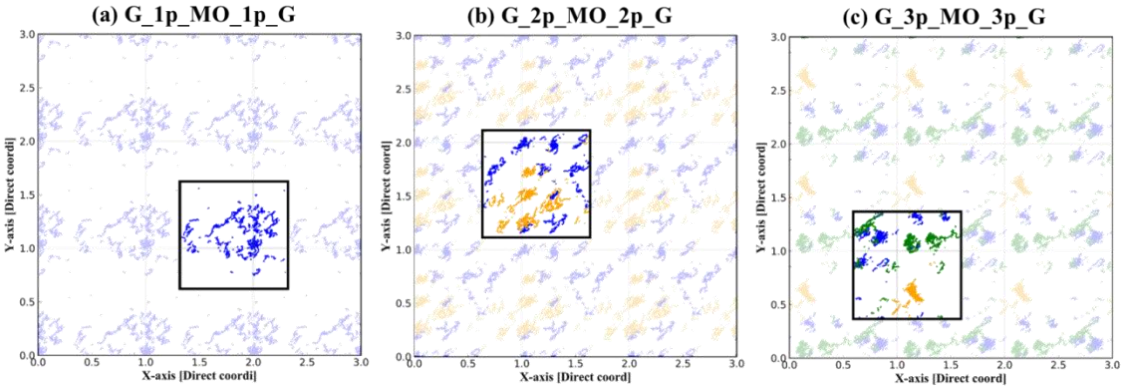

**Fig. 4.** Trajectories of protons at the $Ti_3C_2O_2$-graphene interface as monitored by hydronium O atoms and hydroxyl O atoms on $Ti_3C_2O_2$ surfaces: **(a)** G_1p_MO_1p_G; **(b)** G_2p_MO_2p_G; **(c)** G_3p_MO_3p_G. A 3×3 frame (9 repeating units of the supercell) is shown for a single interface; each color denotes a different proton.

Next, we analyzed proton mobility at the $Ti_3C_2O_2$-$Ti_3C_2O_2$ interface for comparison. In all three proton concentrations, we found that proton prefers to stay in water 99% of time, so proton diffusion is dominated by the in-water proton transfer. From the trajectories of hydronium ions (Fig. 5), we found apparent gaps between different hydronium trajectories; in other words, proton motion is localized and the more protons, the more localized, leading to the decreasing trend of proton diffusivity in the $Ti_3C_2O_2$-$Ti_3C_2O_2$ interface.

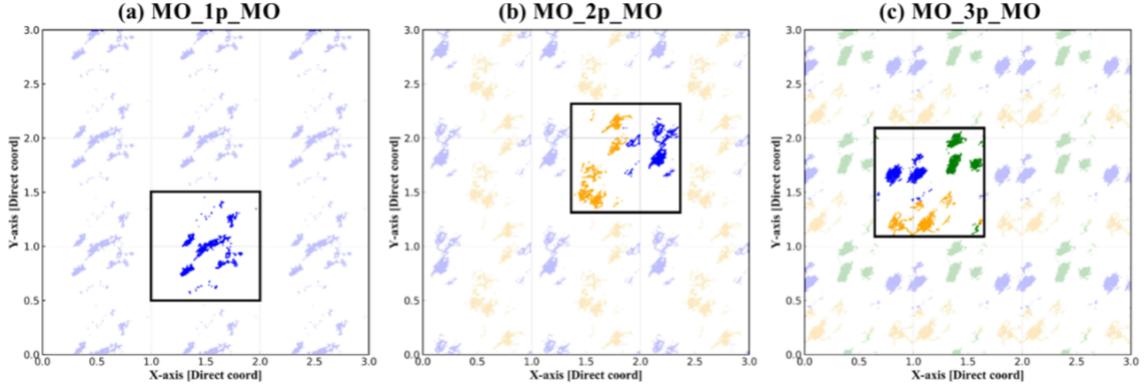

**Fig. 5.** Trajectories of protons at the $Ti_3C_2O_2$-$Ti_3C_2O_2$ interface as monitored by hydronium O atoms and hydroxyl O atoms on $Ti_3C_2O_2$ surfaces: **(a)** MO_1p_MO; **(b)** MO_2p_MO; **(c)** Mo_3p_MO. A 3×3 frame (9 repeating units of the supercell) is shown for a single interface; each color denotes a different proton.

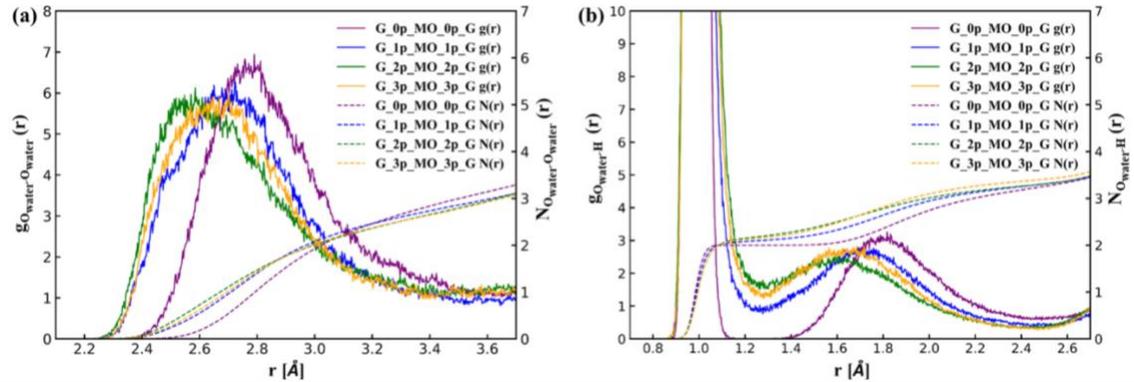

**Fig. 6.** Radial distribution function [g(r), left axis] and coordination number [N(r), right axis] at the $Ti_3C_2O_2$-graphene interface: (a) $O_{water}$-$O_{water}$; (b) $O_{water}$-H.

## C. Water mobility and structure at the interfaces

Comparing proton diffusivity (Fig. 3) and trajectories at the dissimilar $Ti_3C_2O_2$-graphene interface (Fig. 4) and the similar $Ti_3C_2O_2$-$Ti_3C_2O_2$ interface (Fig. 5), the most striking feature is the opposite trend from 1 to 2 protons in the water layer. Two factors may contribute to this different trend in proton mobility in water: (i) water mobility itself; (ii) water structure. We found that water diffusivity in the dissimilar $Ti_3C_2O_2$-graphene interface decreases sharply with proton concentration (Fig. S3), so we can rule out water mobility in accelerating proton mobility at the $Ti_3C_2O_2$-graphene interface.

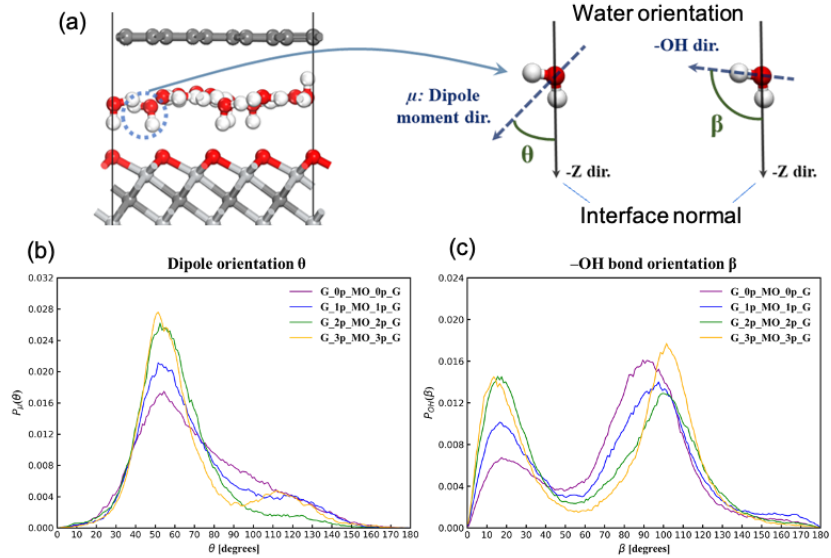

**Fig. 7**. Water orientation at the $Ti_3C_2O_2$-graphene interface with different numbers of interfacial protons: (a) definition of the water dipole (θ) and O-H bond (β) orientations along the interface normal; (b) normalized distribution of the water dipole; (c) normalized distribution of the water OH bonds.

We next turn our attention to water structure at the interface. The radial distribution function (RDF) and coordination number of the $O_{water}$-$O_{water}$ and $O_{water}$-H are plotted in Fig. 6. When changing from G_1p_MO_1p_MO to G_2p_MO_2p_MO, both the $O_{water}$-$O_{water}$ peak [Fig. 6(a), from 2.8 to 2.4 Å] and H-bond peak [Fig. 6(b), from 1.8 to 1.6 Å] shifted left, indicating shortened distances and denser hydrogen-bond networks within the interface. We also investigated the water dipole orientation and OH bond orientation (Fig. 7). One can see that as proton concentration increases, the orientation of the water dipole [Fig. 7(b)] shifts to the region at ~52° and the O-H bonds [Fig. 7(c)] are aligned toward two regions (10°<β<30° and 80°<β<110°). Because the H-O-H angle in water is about 105°, the results in Fig. 7 suggest formation of more directional water structures with one OH bond from water molecules pointing directly toward the $Ti_3C_2O_2$ surface, leading to denser hydrogen-bond network (Fig. 6). This is the reason for enhanced proton mobility in G_2p_MO_2p_MO, as further supported by the analysis below.

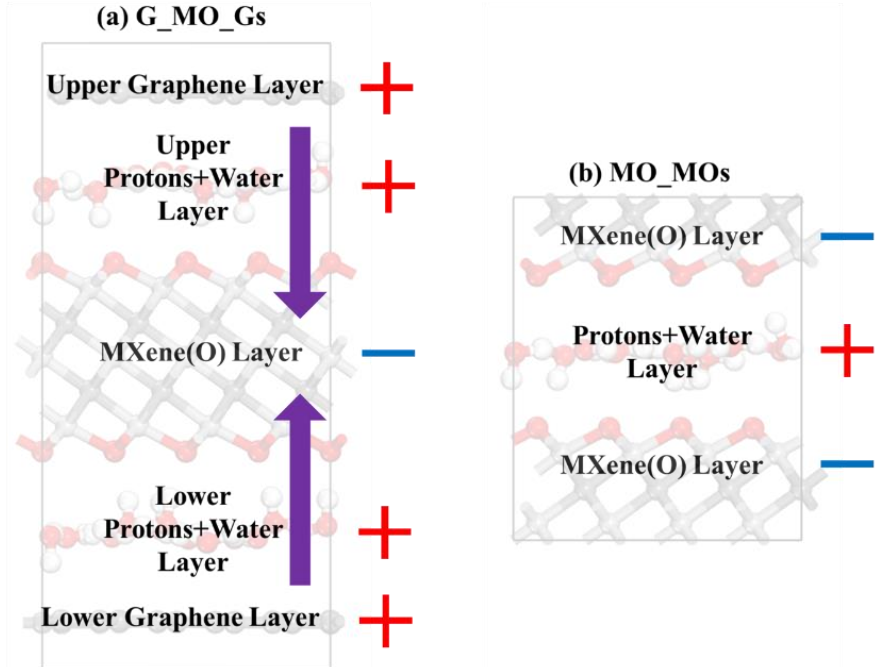

**Fig. 8**. A schematic of the charge distribution (+/- signs) and interfacial electric field (purple arrows): (a) water/hydronium confined in the $Ti_3C_2O_2$-graphene interfaces; (b) water/hydronium confined between $Ti_3C_2O_2$ layers.

**D.     Interfacial polarization and its impact on water structure at the interface**

We think that the interfacial charge transfer and the resulting polarization is the reason behind the denser and more oriented water structure at the $Ti_3C_2O_2$-graphene interface. In the case of MXene/graphene heterostructures without the electrolyte, the charge transfer has already been shown to be from graphene to $Ti_3C_2O_2$.[37] We further confirmed that this is also the case when water/hydronium molecules are presented in the the $Ti_3C_2O_2$-graphene interface from a detailed Bader charge analysis (Fig. S4). In other words, an interfacial electric field directed from graphene to $Ti_3C_2O_2$ across the water/hydronium layer will always be there [Fig. 8(a)], which aligns water molecules. Such field is missing in the similar $Ti_3C_2O_2$-$Ti_3C_2O_2$ interface [Fig. 8(b)]. As more protons are intercalated into the $Ti_3C_2O_2$-graphene interfaces, the greater interfacial electric field across the water/hydronium layer causes stronger interfacial interactions and shorter $O_{water}$-$O_{MXene}$ distances (Fig. S5). This closer interaction together with the aligned O-H bonds pointing toward surface O atoms on $Ti_3C_2O_2$ surfaces enhances proton transfer both in water and across the $Ti_3C_2O_2$-graphene interface. We note that, as done previously, plotting the average electrostatic potential[51,52] and the charge-density variation[53] is a more accurate approach to describe the interfacial polarization, instead of using the partial atomic charges such as the Bader charges due to the certain arbitrariness in partitioning the electrons to atoms; nevertheless, the Bader charges have been shown to correlate well with the interfacial polarization across the heterostructure interface.[37,54] In addition, the interfacial polarization as shown in Fig. 8(a) is not a result of the symmetric cell that we have used, because a similar amount of charge transfer in the same direction was also found at the interface in the asymmetric cell that we tested.

**E.    Further discussion**

Nuclear quantum effects (NQEs) have been shown to be important in describing proton diffusion in confined water,[55,56] but they tend to be very computationally expensive to include. The present AIMD shows a picture of proton diffusion in a monolayer of water confined in the $Ti_3C_2O_2$-graphene interface by treating the proton as a classical particle with the forces derived from DFT within the Born-Oppenheimer approximation. Looking ahead, it would be more desirable to include NQEs for a more accurate description of proton motion in confined water, with efficient approaches such as machine learning.[55]

Reorganization of the hydrogen-bond work has been shown to be important in proton transport in bulk water[57] and auto-protolysis of water.[58] In our systems, the water molecules being just one-monolayer thick are very different from the bulk in that they are interacting very strongly with both graphene and the MXene surface. So the reorganization of the hydrogen bond network is slow by our AIMD timescale. We have previously found that, once there are three water layers in the interface, the proton and water dynamics begins to resemble the bulk.[26] Further simulations on these thicker confined water/proton systems in the $Ti_3C_2O_2$-graphene interface, together with the analysis of the associated reorganization of the hydrogen bond network, are warranted to shed light on how the Grotthuss mechanism[59] might be modulated by the interfacial polarization.

In our $Ti_3C_2O_2$-graphene interface, we have used both an ideal termination for $Ti_3C_2O_2$ and a perfect graphene layer. In real experiments, there will be defects and functional groups due to the chemical processing of the building blocks. For example, -OH and -F groups are often present on the $Ti_3C_2$-based MXenes,[60] while graphene oxides are most used to produce solvent-processable graphene flakes via various reduction pathways, inevitably leaving O-containing functionals and defects on the graphene sheet.[61] Hence, it is highly desirable for future simulations to incorporate those functional groups and defects in their models of the interfaces for confined proton transport, to better guide the experimental efforts.

## IV. Conclusions

Using *ab initio* molecular dynamics, we have examined proton transfer and surface redox chemistry as water-solvated hydronium ions confined in the graphene-$Ti_3C_2O_2$ heterostructure. In comparison with the similar interface of water confined between $Ti_3C_2O_2$ layers, we found that proton redox rate is much higher in the graphene-$Ti_3C_2O_2$ interface, due to the formation of a denser hydrogen-bond network with a preferred orientation of water molecules as a result of the interfacial charge transfer and the resulting electric field. This denser and more directional hydrogen-bond network also leads to an optimal proton concentration where in-water proton mobility reaches maximum in the graphene-$Ti_3C_2O_2$ interface. Our work shows that the dissimilar interface offers greater opportunities in tuning interfacial structure and properties to achieve faster proton transfer and redox in the confined water.

**Supplementary Material**

See the supplementary material for details of diffusivity calculations, Bader charges across the interfaces, the analysis of water-MXene interaction, and coordinates for the simulated systems.

**Acknowledgements**

This research is sponsored by the Fluid Interface Reactions, Structures, and Transport (FIRST) Center, an Energy Frontier Research Center funded by the U.S. Department of Energy (DOE), Office of Science, Office of Basic Energy. This research used resources of the National Energy Research Scientific Computing Center, a DOE Office of Science User Facility supported by the Office of Science of the U.S. Department of Energy under contract no. DE-AC02-05CH11231.

**Data availability**

The data that support the findings of this study are available from the corresponding author upon reasonable request.

# Proton dynamics in water confined at the interface of the graphene-MXene heterostructure

Lihua Xu,[1] Tao Wu,[2] Paul R. C. Kent,[3,4] and De-en Jiang[1,2,*]

[1]Department of Chemical and Environmental Engineering, University of California, Riverside, California, 92521, USA

[2]Department of Chemistry, University of California, Riverside, California, 92521, USA

[3]Computational Sciences and Engineering Division, Oak Ridge National Laboratory, Oak Ridge, Tennessee 37831, USA

[4]Center for Nanophase Materials Sciences, Oak Ridge National Laboratory, Oak Ridge, Tennessee 37831, USA

*E-mail: djiang@ucr.edu

1. **Calculation details about proton diffusivity**

The time-dependent mean square displacement (MSD) of the proton-bonded $O_0$ were estimated from the following expression:

$$MSD(t) = \langle r^2(\Delta t) \rangle = \langle (r_i(\Delta t + t_0) - r_i(t_0))^2 \rangle$$
$$= \frac{1}{N} \sum_{i=1}^{N} (r_i(\Delta t + t_0) - r_i(t_0))^2,$$

where $r_i(\Delta t + t_0)$ and $r_i(t_0)$ represent the position of particle $i$ at time $(\Delta t + t_0)$ and the reference position at $t_0$, respectively. $N$ is the total number of snapshots. The diffusion coefficient (D) is estimated by the Einstein relation—the slope of the MSD divided by $n$. In our study, the slop is obtained by fitting the virtual linear region from the MSD curves.

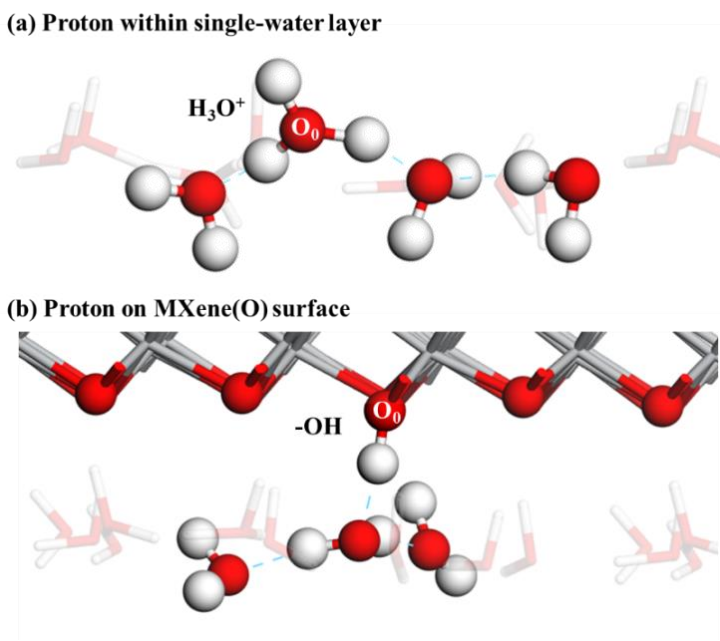

**Fig. S1.** The identity of the proton-bonded $O_0$ inside **(a)** $H_3O^+$ within single water-layer, or **(b)** -OH group on MXene surface.

## 2. Total/decomposed proton diffusion coefficients

The total MSD as a function of time for the proton-bonded $O_0$ is shown in **Fig. S2(a)**. To evaluate the contributions to the total diffusion coefficients, the decomposed XY-plane ($n = 4$ for 2D system) and Z-direction ($n = 2$ for 1D system) (**Fig. S2(b)**) diffusion coefficient are obtained by projecting the proton-bonded O positions to the relative direction in MSD. Since single water-layer is applied at each interface, the XY-plane proton diffusion is mainly due to proton in-water transfer, while the Z-direction diffusion corresponds to proton jumping between water and MXene surface (namely surface redox reactions). As a result, the XY-plane diffusion contributed primarily to the proton diffusion in all MO_MOs. In G_MO_Gs, the incipient increase of the diffusion coefficient should be attributed to the faster proton in-water transfer, and the latter decreased value is due to the diffusion restrictions derived from the more frequent proton surface-redox behavior.

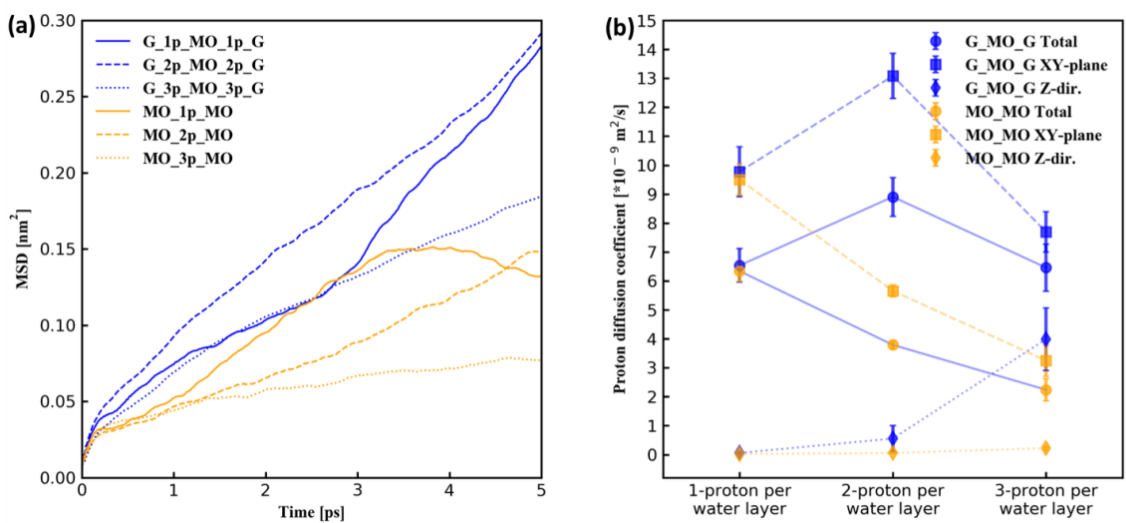

**Fig. S2.** Proton diffusion within the interface: **(a)** Total mean-squared displacement (MSD) as a function of time for $O_0$; **(b)** Total and decomposed (XY-plane and Z-dir.) proton diffusion coefficients for different systems.

### 3. Total/decomposed $O_{water}$ diffusion coefficients

The diffusion coefficients are found to be in a decreasing trend for O in water layer ($O_{water}$) with increased intercalating proton concentration, indicating that water molecules become more motionless (**Fig. S3**). Here, we included the $H_3O^+$ species in our calculations because they influence the water dynamic behavior dramatically within the single water-layer.

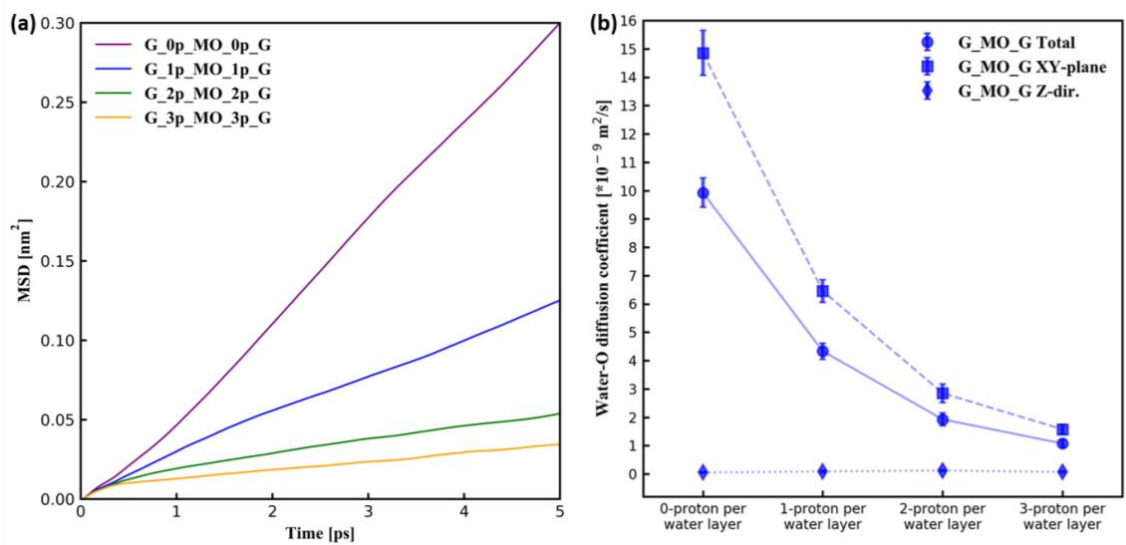

**Fig. S3.** The $O_{water}$ diffusion within the interface: **(a)** Total mean-squared displacement (MSD) as a function of time for $O_{water}$; **(b)** Total and decomposed (XY-plane and Z-dir.) $O_{water}$ diffusion coefficients.

## 4. Bader charge analysis for at the interfaces

We calculated the Bader charge for different building blocks of 10 randomly selected snapshots for all systems under investigation. The averaged Bader charge values with small standard deviations were presented in **Fig. S4**. Accompanied by the increased intercalating proton concentration, for G_MO_Gs, fraction of electrons goes from interfaces ("Upper/Lower Protons+Water Layer") to graphene and another fraction goes to MXene, resulting in a more negatively charged MXene and more positively charged interface. The slight reduced positive charge in graphene does not impact the presence of the interfacial electric field with the direction from graphene to MXene. Contrary to G_MO_G, no interfacial electric field appears within the similar interface in MO_MOs but the MXene still becomes more negatively charged due to the interfacial electron transfer.

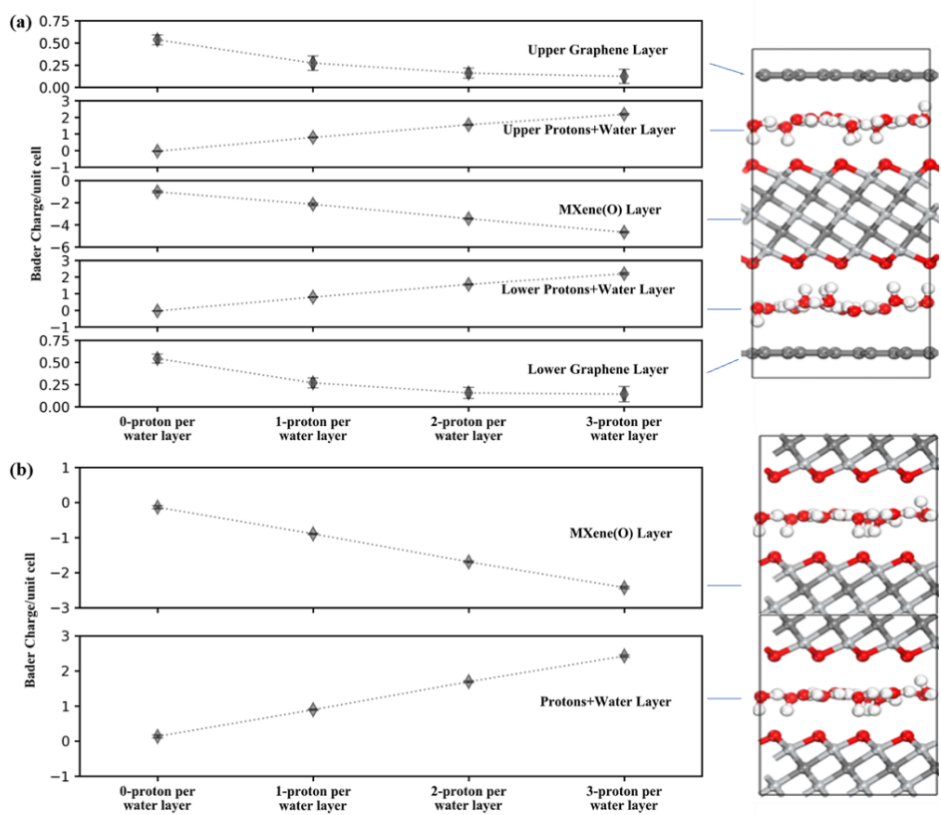

**Fig. S4.** The averaged Bader charge (from 10 randomly selected snapshots) for each part of the interface of the unit cell: **(a)** G_MO_G; **(b)** MO_MO.

## 5. Radial distribution function (RDF) of $O_{water}$-$O_{MXene}$

**Fig. S5** displays the radial distribution function (RDF) g(r) and coordination number N(r) of the distance between the O from water ($O_{water}$) and the surface-O from MXene ($O_{MXene}$) for different G_MO_Gs. The left-shift peak in terms of additional intercalating proton suggests a reduced $O_{water}$-$O_{MXene}$ distance. Hence, the water layer comes more closer to the MXene surface, leading to a higher possibility for protons to transfer between water and MXene surface.

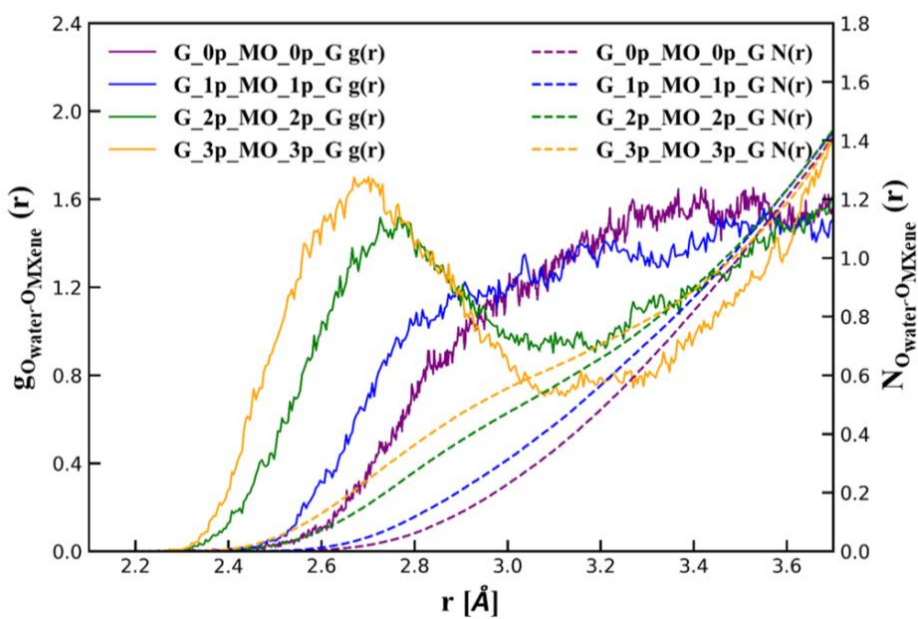

**Fig. S5**. The radial distribution function (RDF) g(r) and coordination number N(r) of $O_{water}$-$O_{MXene}$ for G_MO_Gs.